\documentclass[preprint,showkeys,showpacs,preprintnumbers,amsmath,amssymb]{revtex4}

\begin{document}


\title{Continuous distribution of frequencies and deformed dispersion relations}
\author{Abel Camacho}
\email{acq@xanum.uam.mx} \affiliation{Departamento de F\'{\i}sica,
Universidad Aut\'onoma Metropolitana--Iztapalapa\\
Apartado Postal 55--534, C.P. 09340, M\'exico, D.F., M\'exico.}

\date{\today}

\begin{abstract}
The possibilities that, in the realm of the detection of the
so--called deformed dispersion relation, a light source with a
continuous distribution of frequencies offers is discussed. It
will be proved that the presence of finite coherence length
entails the emergence of a new term in the interference pattern.
This is a novel trait, which renders a new possibility in the
quest for bounds associated with these deformed dispersion
relations.
\end{abstract}
\maketitle

\section{Introduction}

In the extant literature many ideas can be found whose goal is a
quantum theory of gravity. Some of them entail the modification of
the dispersion relation \cite{[1]}. These approaches appear in
different realms, for instance, quantum--gravity approaches based
upon non--commutative geometry \cite{[2], [3]}, or loop--quantum
gravity models \cite{[4], [5]}, etc. In them Lorentz symmetry
becomes only an approximation for quantum space \cite{[6], [7]}.
Though concerning the feasibility of detecting this kind of
changes some advance has already been done, it must be strongly
underlined that these experimental proposals have always embraced
light sources containing a discrete number of frequencies
\cite{[1], [8], [9]}, where each one of these beams has vanishing
frequency width.

The last condition has always been introduced as an idealization,
and thereby, supposedly, an advantage in the analysis of the
situation has always been present. In the present work, the case
of a beam with non--vanishing width frequency will be considered,
i.e., a finite coherence length will be an important trait of the
experimental idea.

It will be shown that this novel feature not only does not entail
a more cumbersome situation (as could be guessed since we lose an
{\it idealization}), but it yields a new term in the interference
pattern, absent in the previous analyses, the one hinges upon the
ratio between Planck's length and the longitudinal coherence of
the corresponding beam. This fact implies the presence of an
additional possibility in the detection of this kind of quantum
gra\-vity effects.
\bigskip
\bigskip

\section{Finite longitudinal coherence length and quantum--gravity effects}

Several quantum--gravity models predict a modified dispersion
relation \cite{[1],[2], [3], [4], [5]}, the one can be
characterized, from a phenomenologically through corrections
hinging upon Planck's length, i.e., $l_p$

\begin{equation}
E^2 = p^2\Bigl[1 - \alpha\Bigl(El_p\Bigr)^n\Bigr].
\label{Disprel1}
\end{equation}

Here $\alpha$ is a coefficient, usually of order 1 and whose
precise value depends upon the considered quantum--gravity model,
and $n$, the lowest power in Planck's length leading to a
non--vanishing contribution, is also model dependent. In ordinary
units (\ref{Disprel1}) becomes

\begin{equation}
E^2 = p^2c^2\Bigl[1 -
\alpha\Bigl(E\sqrt{G/(c^5\hbar)}\Bigr)^n\Bigr]. \label{Disprel2}
\end{equation}

Recalling that

\begin{equation}
p =\hbar k. \label{Mom1}
\end{equation}

We conclude

\begin{equation}
k =\frac{E/(c\hbar)}{\Bigl[1 -
\alpha\Bigl(E\sqrt{G/(c^5\hbar)}\Bigr)^n\Bigr]^{1/2}}. \label{k1}
\end{equation}

The modifications are quite small, then the following expansion is
justified

\begin{equation}
k =\frac{E}{c\hbar}\Bigl[1 +
\frac{\alpha}{2}\Bigl(E\sqrt{G/(c^5\hbar)}\Bigr)^n +
\frac{3}{8}\alpha^2\Bigl(E\sqrt{G/(c^5\hbar)}\Bigr)^{2n}+...\Bigr].
\label{k2}
\end{equation}

Let us now consider the so--called spectral distribution function,
$P(\omega)$, defined as

\begin{equation}
P(\omega) = I(\omega)/I_0. \label{power}
\end{equation}

Here we have that $I_0$ denotes the integrated intensity of our
source

\begin{equation}
I_0= \int_0^{\infty}I(\omega)d\omega. \label{int1}
\end{equation}

In these two last expressions it has been, implicitly, assumed
that our sources have attached, as corresponding parameters,
integrable functions. If a more general situation is to be
analyzed, then we may resort to the definition of the spectral
distribution function in terms of the density of photons in the
differential interval $d^3k$, which in turn can be deduced, for a
thermal equilibrium source, from the mean photon number
$<n_{\vec{k},s}>$ ($\vec{k}$ denotes the direction of propagation,
while $s$ is the polarization for each $\vec{k}$) \cite{[10]}

\begin{equation}
<n_{\vec{k},s}> = [\exp{(\hbar\omega/k_BT)} -1]^{-1}\label{SDF}.
\end{equation}

Here $k_B$ is Boltzmann's constant, and $T$ the temperature.

Let us restrict ourselves to the case in which the corresponding
functions are integrable, then the interference pattern to be
detected in a Michelson interferometer, assuming that the detector
cannot respond to the beat frequencies produced on the surface of
the detector by two waves reads \cite{[10]}

\begin{equation}
I = I_0\Bigl[1 + \gamma(\tau)\Bigr], \label{Int2}
\end{equation}

where

\begin{equation}
\gamma(\tau)= \int_0^{\infty}P(\omega)\cos(\omega\tau)d\omega
\label{int3},
\end{equation}

is usually called the degree of coherence, and

\begin{equation}
\tau= 2d/c'\label{time1},
\end{equation}

is the so--called retardation time. In other words, it is the
difference in the arrival time (at the detector) ori\-ginated in
our Michelson interferometer due to the fact that in this kind of
device there are two arms with an optical path length showing a
difference of $d$ \cite{[10]}. In our case we have a more
complicated situation since now the speed, say $c'$, has a
non--trivial energy dependence \cite{[9]}

\begin{equation}
c' = c\Bigl[1
-\alpha\Bigl(E\sqrt{G/(c^5\hbar)}\Bigr)^n\Bigr]^{1/2}\Bigl[1
+\frac{\alpha n\Bigl(E\sqrt{G/(c^5\hbar)}\Bigr)^n}{1-\alpha
\Bigl(E\sqrt{G/(c^5\hbar)}\Bigr)^n}\Bigr]^{-1} \label{speed1}.
\end{equation}

Clearly. setting $\alpha= 0$ in this last equation renders the
usual speed of light, namely, an expression independent of the
energy of the beam.

In the following we will consider the case in which $n = 1$, this
situation can be easily found in the realm of non--commutative
geometry \cite{[1]}. Under this restriction we have that

\begin{equation}
\omega\tau = \frac{2dE}{c\hbar} \frac{1
+(\alpha/2)E\sqrt{G/(c^5\hbar)}}{\Bigl[1
-\alpha\Bigl(E\sqrt{G/(c^5\hbar)}\Bigr)\Bigr]^{3/2}}
\label{prod1}.
\end{equation}
\bigskip
\bigskip

\section{Finite coherence length and deformed dispersion relations}

Clearly, the analysis of the feasibility of the present proposal
requires some particular expression for the spectral distribution
function, see (\ref{power}). At this point two cases will be
analyzed.
\bigskip
\bigskip

\subsection{Gas at low pressure}

For a gas at low pressure, in thermal equilibrium, the spectral
distribution function has a Gaussian form \cite{[10]}, in which
the corresponding linewidth, say $\tau_d$, is related to the
physical properties of the gas by the following expression

\begin{equation}
\tau^2_d = \frac{c^2m}{8\omega^2_0\kappa T}\label{linewidth},
\end{equation}

\begin{equation}
P(\omega) =
\tau_d\sqrt{\pi}\exp\Bigl[-(\omega-\omega_0)^2(\tau_d/2)^2\Bigr]
\label{power1}.
\end{equation}

Here $\omega_0$ denotes the center of the frequency, $m$ the
molecules mass, $T$ the temperature, and finally, $\kappa$
Boltzmann constant.

After a messy integration process we deduce that the corresponding
degree of coherence reads

\begin{eqnarray}
\gamma(d)=\sqrt{\pi}\exp\Bigl[-\frac{31}{32}\Bigl(\frac{z_0\tau_d}{2b\hbar}\Bigr)^2
-\frac{1}{2}\Bigl(\frac{b\hbar d}{\tau_d}\Bigr)^2\Bigr]\nonumber\\
\times\Bigl[\frac{1}{\sqrt{8}}\Bigl(e^{z_0d/16}D_{(-1)}(\sqrt{2}[\frac{z_0d}{8b\hbar}
+i\frac{b\hbar d}{\tau_d}])\nonumber\\
+e^{-z_0d/16}D_{(-1)}(\sqrt{2}[\frac{z_0d}{8b\hbar}
-i\frac{b\hbar d}{\tau_d}]\Bigr)\nonumber\\
+6\alpha\frac{l_p}{l_d}\Bigl(e^{z_0d/16}D_{(-2)}(\sqrt{2}[\frac{z_0d}{8b\hbar}
+i\frac{b\hbar d}{\tau_d}])\nonumber\\
+e^{-z_0d/16}D_{(-2)}(\sqrt{2}[\frac{z_0d}{8b\hbar} -i\frac{b\hbar
d}{\tau_d}])\Bigr)\Bigr]\label{Degree1}.
\end{eqnarray}

In the last expression $l_d= c\tau_d$ denotes the cohe\-rence
length associated to our beam, see \cite{[10]}, whereas, $l_p$ is
Planck's length, and, $b=2/(c\hbar)$, $z_0
=\frac{2E_0}{c\hbar}\Bigl[1 +2\alpha
E_0\sqrt{G/(c^5\hbar)}\Bigr]$. In addition, $D_{(-1)}$ and
$D_{(-2)}$ are the so--called parabolic cylinder functions
\cite{[11]}.
\bigskip
\bigskip

\subsection{Gas at high pressure}

For a gas at high pressure the spectral distribution function has
a Lorentzian form \cite{[10]}

\begin{equation}
P(\omega) =
\frac{B}{arctg(-\omega_0/B)}\frac{1}{(\omega-\omega_0)^2 + B^2}
\label{power2}.
\end{equation}

In this case the coherence time is $\tau_d = B^{-1}$, and in
consequence the coherence length reads $l_d = c/B$.

\begin{eqnarray}
\gamma(d)= \frac{\pi\cos(z_0d)}{2arctg(-\omega_0/B)}\Bigl[1
-3\alpha l_pz_0\Bigr]e^{-2d/l_d}\nonumber\\
-\frac{\sin(z_0d)}{2arctg(-\omega_0/B)}\Bigl[1
-3\alpha l_pz_0\Bigr]\nonumber\\
\times\Bigl(e^{-2d/l_d}\hat{E}_i(2d/l_d) -
e^{2dl_d}E_i(-2d/l_d)\Bigr)\nonumber\\
+  3\frac{l_p}{l_d}\frac{\alpha
}{arctg(-\omega_0/B)}\nonumber\\
\times\Bigl\{\cos(z_0d)\Bigl(e^{-2d/l_d}\hat{E}_i(2d/l_d) +
e^{2dl_d}E_i(-2d/l_d)\Bigr)\nonumber\\
 + \pi e^{-2d/l_d}\sin(z_0d)
\Bigr\}\label{Degree2}.
\end{eqnarray}

In (\ref{Degree2}) the following definition has been introduced

\begin{equation}
z_0 = \frac{2E_0}{c\hbar}\Bigl[1 + 2\alpha
E_0\sqrt{G/(c^5\hbar)}\Bigr]\label{Definition1}.
\end{equation}

Where $E_0$ is the energy related to $\omega_0$, and $\hat{E}_i$,
$E_i$ are the so--called related exponential integral and
exponential integral function \cite{[11]}, respectively. A careful
analysis of the effect of a small coherence length can be done
noting that the so--called parabolic cylinder functions are
closely related to the Hermite polynomials \cite{[11]}, and in
consequence the degree of coherence function becomes, in terms of
the coherence length, a polynomial, the one is quite easy to
analyze for the aforementioned situation.
\bigskip
\bigskip

\section{Conclusions}

A fleeting glimpse at (\ref{Degree1}) and (\ref{Degree2}) shows us
that a new term, absent in all previous work in this context
\cite{[1], [9]}, emerges, namely, the one hinging upon $l_p/l_d$.
This fact that can readily be explained. Indeed, in the
aforementioned proposals the coherence length involved has always
been infinite, since they consider monocromatic beams. In other
words, the possi\-bility of detecting a modified dispersion
relation seems to improve if the involved beam has a finite
coherence length.

In order to fathom better some of the consequences that the
presence of a deformed dispersion relation could have upon an
interference experiment, let us look at the the case of a gas at
low pressure. It is readily seen that in this situation the
coherence length appears in the extra terms not only as part of
the argument of the so called cylinder parabolic functions, see
(\ref{Degree1}). This novel dependence of the degree of coherence
function upon $l_d$ is absent in the usual situation, i.e., when
$\alpha =0$, and implies that the way in which $\gamma(d)$ changes
as a function of $l_d$ could be exploited to confront the
prediction of the present model against measurement readouts. As
for the case of a gas at high pressure it is noteworthy to mention
that the situation $\alpha\not =0$ entails the presence of the
same kind of modifications in $\gamma(d)$, as in the previous
analysis, and therefore the conclusions are also valid.

Under ideal circumstances, and by this phrase we mean $l_p\sim
l_d$, the order of magnitude of the new term, see (\ref{Degree1})
and (\ref{Degree2}), will have the same order of magnitude as the
usual contribution. In a more realistic situation, the order of
magnitude depends, crucially, upon the value of $l_p/l_d$. As the
coherence length gets closer to Planck's length the role of the
deformation becomes more relevant in the determination of
$\gamma(d)$.

In our results we have that the presence of a deformed dispersion
relation embraces not only the usual contribution, but a new one,
which is a little bit puzzling, since it predicts interference
even with a very small coherence length. The answer to this
paradox stems from the fact that here we are dealing not with the
usual electrodynamics, but with a modification of it, and
therefore we shall expect the emergence of new effects. In other
words, the possibility of having interference, even with a very
small coherence length, is a novel trait appearing as a
consequence of the effects of quantum gravity upon the usual
electromagnetic radiation.

 Let us analyze a little deeper the case of
a gas at low temperature. Demanding $l_p/l_d\sim 1$, and accepting
the validity of the Theorem of Equipartition of Energy \cite{[12]}
we have

\begin{equation}
<v^2> \sim c^2\frac{E_0}{E_p}\label{squarespeed1}.
\end{equation}

Here $E_0$ denotes the energy associated to $\omega_0$, see
(\ref{power1}). If we consider an optical transition, i.e.,
$E_0\sim 1 e-V$, then

\begin{equation}
\sqrt{<v^2>} \sim 10^{-6}cm/s\label{squarespeed2}.
\end{equation}

Resorting, once again, to the Theorem of Equipartition of Energy,
and assuming particles with a mass of $m\sim 10^{-24}g$, we deduce
that the gas must be at a temperature of

\begin{equation}
T \sim 10^{-14} K\label{temperature}.
\end{equation}

At this order of magnitude of the temperature Boltzmann statistics
is not valid \cite{[12]}, and in consequence one of our
assumptions, i.e., the Theorem of Equipartition of Energy cannot
be employed. The analysis of more promising situations, for
instance, gamma ray bursts unavoidably requires the knowledge of
the corresponding spectral distribution function.

Summing up, it has been shown, resorting to two di\-fferent
situations for the spectral distribution function, namely Gaussian
and Lorentzian cases, that the presence of a finite longitudinal
coherence length for the spectral distribution function entails
the emergence of a new term in the interference pattern the one
depends upon the ratio between Planck's length and the coherence
length, i.e., $l_p/l_d$, the one is absent in the usual analysis,
in which $l_d\rightarrow\infty$. The main lesson to be elicited is
the fact that an ideal beam (as those considered in previous
proposals, see \cite{[9]}, for instance) will not render an
interference pattern containing those terms proportional to
$l_p/l_d$, and hence the possibility of detecting deformed
dispersion relations could be, in the context of them, more
difficult to do.

Finally, let us address the issue of the possibilities of the
present technology, if one tries to carry out this kind of
experiment in a laboratory. Clearly, (\ref{temperature}) shows us
that a gas in the classical limit will not be a solution to the
experimental problem. Now, since we require $l_p/l_d$ not much
smaller than $1$, then we need a coherence time, $\tau_d$, such
that it fulfills the condition $l_p/\tau_d\sim c$, and this
implies $\tau_d\sim 10^{43}s$. Considering that, in a very rough
approximation, the coherence length of a He--Ne laser reads
$l_d\sim 10$cm, then the corresponding coherence time is
$\tau_d\sim 10^{-9}s$, quite far away from our requirement. These
last arguments discard any experiment carried out with a laser,
which, by the way, is the human--made device with the largest
coherence time. Hence, it seems that we must consider
astrophysical sources, for instance, Gamma Ray Busts \cite{[13]},
in order to have this kind of coherence times.

\begin{acknowledgments}
This research was supported by CONACYT Grant 42191--F. The author
would like to thank A.A. Cuevas--Sosa and A. Mac\'{\i}as for
useful discussions and literature hints. Additionally, it has to
be acknowledged the enrichment that the comments and remarks of
the referee have brought to the present work.
\end{acknowledgments}

\end{document}